\documentclass[prb,twocolumn,groupedaddress,superscriptaddress]{revtex4}
\usepackage{eurosym}
\usepackage{amsfonts}
\usepackage{amsmath}
\usepackage{amssymb}
\usepackage{graphicx}
\usepackage{float}

\begin{document}

\title{Temperature changes of the Fe$_{8}$ molecular magnet during its spin reversal process}

\author{Maayan Yaari}
\affiliation{Department of Physics, Technion - Israel Institute of Technology, Haifa, 32000, Israel}

\author{Amit Keren}
\affiliation{Department of Physics, Technion - Israel Institute of Technology, Haifa, 32000, Israel}

\date{\today }

\begin{abstract}

Tunneling of the spins in the Fe$_8$ molecular magnet from a metastable ground state to an excited state is accompanied by a decay of these spins to the global ground state, and an increase of the crystal temperature. We measured this temperature using two thermometers, one strongly coupled and the other weakly coupled to the thermal bath. We found that the temperature increases to no greater than $2.2$~K. This upper limit agrees with the flame temperature derived from deflagration theory and previous measurements. In light of this temperature increase we re-examine the Landau, Zener and Stuckelberg (LZS) theory of spin tunneling in large Fe$_8$ crystals.

\end{abstract}

\maketitle

\footnotetext[1]{Correspondence should be addressed to A.K. (email: keren@physics.technion.ac.il)}

The Fe$_8$ single molecular magnet is an exciting system to study since its dynamics are fully quantum mechanical below a temperature of $400$~mK~\cite{Sangregorio1997}. This molecule has a spin of $S=10$, and accounting for the “crystal-field” together with the spin-orbit interaction, it is governed by the spin Hamiltonian~\cite{nanomagnets}:
\begin{equation}
 \mathcal{H} = DS_z^2 +\mathcal{H_\perp} +g\mu_B\vec{S}\cdot\vec{H}
  \label{eq:Ham}
\end{equation}
where the dominating $S_z^2$ term with $D=-0.295$~K gives rise to an anisotropy barrier \cite{Barra2000,Caciuffo1998,Mukhin2001}. The $\mathcal{H_\perp}$ term is responsible for the mixing of spin states and tunneling between them.  
The Zeeman term removes the degeneracy between $S_z = \pm m$ and allows the spins of all molecules to align at sufficiently low temperatures. Upon sweeping of the magnetic field from $H_0$ to $-H_0$, the samples magnetization versus field curve exhibits a staircase hysteresis loop~\cite{LeviantDeflagration}. This is attributed to quantum tunneling between magnetization states, which is only allowed for discrete 'matching fields' corresponding to level crossings~\cite{Wernsdorfer2000,Caneschi1999,WernsdorferSessoli1999,EPR2002}. The matching fields for transitions between the states $m$ to $m'$ are given by:

\begin {equation}
H_{n}=Dn/g\mu_B\simeq 0.225[\text{T}]n
\label{eq:matchfield}
\end{equation}
where $n=m+m'$~\cite{Wernsdorfer2000}. 

Due to $\mathcal{H_\perp}$ the level crossing is in fact an avoided crossing with a tunnel splitting $\Delta_{mm'}$ between the $m$ and $m'$ levels. According to the Landau, Zener and Stuckelberg [LZS] solution~\cite{ Landau1932, Zener1932,Stuckelberg1932} of the time dependent Schr\"odinger equation for a multistate system, the probability for transition between two states, when the external field is in the vicinity of a matching field is given by:

\begin{equation}
 {P_{mm'}=1-exp(\frac{-\pi\Delta_{mm'}^2}{2 g\mu_B (m-m')\alpha_B})},
 \label{eq:LZ}
\end{equation}
where for an isolated spin

\begin{equation}
\alpha_B=\alpha_H=\mu_0 \frac{dH_z}{dt}.
\end{equation}

However, upon sweeping the field at low temperatures, the transitions occur between a metastable spin state (say $m=-10$), and either a ground state ($m=10$) or an excited state (e.g. $m=9$). We name these transitions according to their $n$ value. For $n \geq 1$ these transitions are accompanied by a decay to the stable ground state (e.g. $9 \rightarrow 10$), leading to an energy release and a temperature increase. According to the spin Hamiltonian, the lowest energy difference between a metastable state and the ground state is approximately $5$~K. Therefore, the decay should lead to substantial heating of the crystal and affect the transition rates. In these circumstances the LZS formula will not be applicable for all transitions other then $n=0$, namely $\pm 10$ to $\mp 10$. Therefore, to properly account for the tunneling probability of general molecular magnets embedded in a crystal and Fe$_8$ in particular, it is essential to determine how hot the crystal gets after a tunneling event that is followed by an energy release. This is the main objective of the experiment reported here. 

\begin{figure}[tbph]
	\begin{center}
		\includegraphics[trim=0cm 0cm 0cm 0cm ,clip=true,width=8cm]{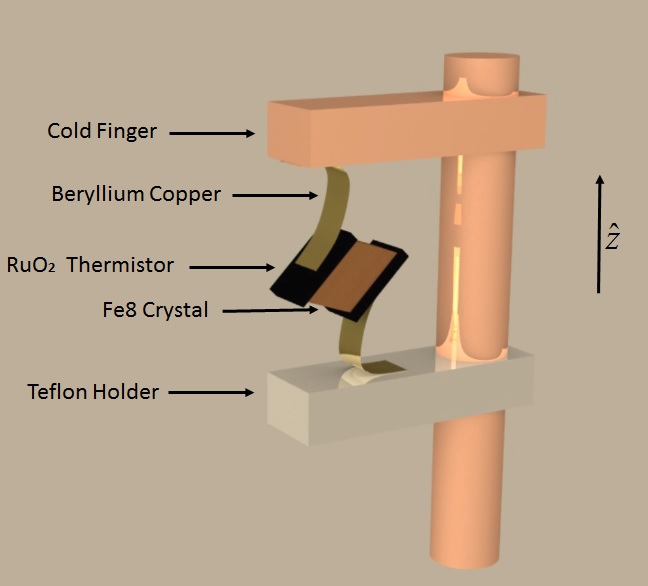}
	\end{center}
	\caption{The experimental setup which is mounted on a cold finger and connected to the $^3$He refrigerator. This setup is in the center of a magnet with the field pointing along the crystal $z$ direction. }
	  \label{setup}
\end{figure}

Our experiment is done using a sorption pumped $^3$He Oxford Instruments refrigerator. For each sample two RuO$_2$ resistance thermometers (thermistors) are glued to the sample with super glue. One of the thermistors is anchored to the refrigerator cold-finger with a copper beryllium spring; we refer to it as the cold thermistor since it is strongly coupled to the cold finger. The other thermistor is anchored to a teflon bar using a similar spring. The teflon bar, in turn, is attached to the cold refrigerator finger; this is the hot thermistor since it is weakly coupled to the cold finger and is expected to warm up more upon energy release. Before the energy release by the molecular magnets both thermistors are at the same temperature. The springs are essential since the facets of the Fe$_8$ crystal are not perpendicular to the $z$ direction of the molecules. The springs also allow for thermal shrinking of the apparatus upon cooling without breaking the crystal. The RuO$_2$ thermistors are not sensitive to magnetic fields. The apparatus is depicted in Fig.~\ref{setup}.

The thermistors resistivity is measured using the four wire method; two wires for current and two for voltage. However, in order to check if heat leaks through these cryogenic wires we modified the wiring between different measurements. Sometimes we connected all four wires directly to the thermistor, and sometimes we used only two wires, which were split into four outside of the cryostat. We found that the wiring method has no impact on the conclusions of our work. We calibrated the thermistors against the built-in thermometer of the $^3$He refrigerator while slowly cooling it to base temperature.
 
We measured six crystals, which are quite different from one another in shape. Consequently their demagnetization factors and internal fields are not identical. This leads to variations in their behavior. Nevertheless, each crystal shows reproducible data when repeating the sweeps and when reversing the sweep direction. Here we present data from two crystals with fundamentally different behavior. 

\begin{figure}[tbph]
	\begin{center}
		\includegraphics[trim=0cm 0cm 0cm 0cm ,clip=true,width=8cm]{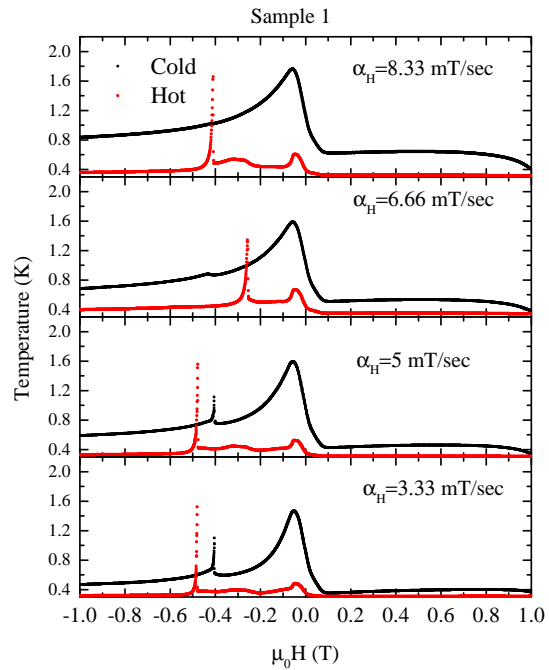}
	\end{center}
	\caption {Temperature versus magnetic field for sample 1, using a four-point probe measurement to determine the resistivity of the thermistor. Each panel shows results for a different magnetic field sweep rate $\alpha_H$.}
	\label{fig:sample3}
\end{figure}

In Fig.~\ref{fig:sample3} we show the sample temperature as recorded by both thermistors while sweeping the magnetic field from positive to negative at four different sweep rates. As the sweep begins the cold thermistor immediately becomes hotter despite its better thermal coupling to the refrigerator. This is due to eddy currents in the copper cold finger and spring. The hot thermistor is hardly affected by the sweep at first. As the field approaches zero, there is a temperature rise. This is believed to be due to the superconducting transition of lead in the soldering material as the field crosses the lead $H_{c1}$. Once the field crosses over to the negative side, the quantum nature of the molecule becomes obvious especially, at the highest sweep rate ($\alpha_H=8.33$~mT/sec) and for the hot thermistor. In this case we see a clear broad temperature increase that starts at $\mu_0 H=-0.2$~T ($n=1$), and a spike in the temperature at $\mu_0 H=-0.42$~T ($n=2$) with a tail towards higher fields. No tunneling events are noticed in the cold thermistor for the highest sweep rate.

At a lower sweep rate of ($\alpha_H=6.66$~mT/sec) we associate the spike at $\mu_0 H=-0.28$~T with the $n=1$ transition. However due to the slow response of the hot thermistor it appears at a slightly higher field. In this case the cold thermistor begins to show some temperature increase at $\mu_0 H=-0.4$~T ($n=2$). 

As we lower the sweep rate further to $5$~mT/sec and then to $3.33$~mT/sec the response of the cold thermistor at $\mu_0 H=-0.4$~T becomes stronger and the spike is again associated with the $n=2$ transition. It is not clear to us why in $\alpha_H=6.66$~mT/sec the largest temperature increase is at $n=1$ and in the other three cases it is at $n=2$. In any case, at the two lowest sweep rates a clear temperature increase is detected in both thermistors. Both spikes are associated with the $n=2$ transition, but there is a field (hence time) delay between them. The delay in terms of field difference is independent of the sweep rate, but of course there is a delay in time as $\alpha_H$ is varied. This phenomenon is intriguing and we lack an explanation for it. Finally, for all sweep rates the temperature does not exceed $2$~K in either of the thermistors.
 
\begin{figure}[tbph]
	\begin{center}
		\includegraphics[trim=0cm 0cm 0cm 0cm ,clip=true,width=8cm]{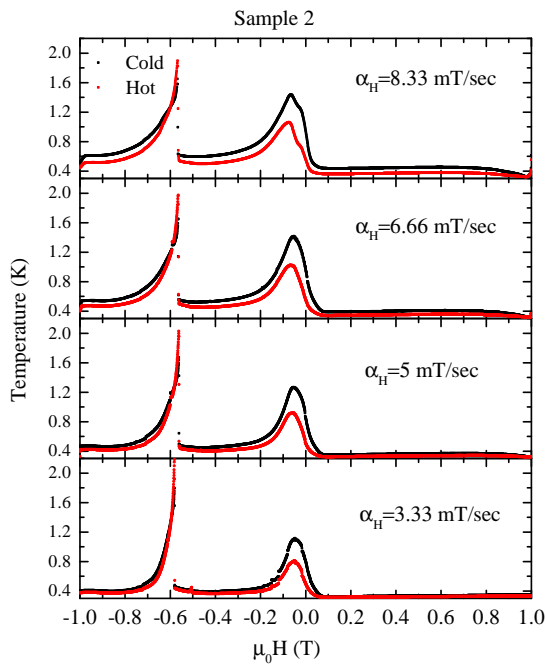}
	\end{center}
	\caption{Temperature versus magnetic field for sample 2, using a 2-point probe measurement to determine the resistivity of the thermistor. Each panel shows results for a different magnetic field sweep rate $\alpha_H$.
	}
	\label{fig:sample2}
\end{figure}

Figure \ref{fig:sample2} represents a special sample where the only tunneling event detected is at $n=3$ for all four sweep rates. In this case both thermistors warm up equally, exactly at the same field (or time). This suggests that all the spins in the sample flip together. For this sample, after the tunneling, the spins decay from the $ \left|m\right| =7 $ excited state to the ground state. Each spin releases an energy of $15$~K. This is approximately three times bigger than the energy release for $n=1$ and approximately 1.5 times bigger than for $n=2$. Yet, the temperature of the sample barely reaches $2$~K. In fact, among all the samples we measured the temperature never reached $2.2$~K. Therefore we conclude that for samples cooled to temperatures of 300 mK or less, where all $n$-states are detected upon sweeping of an external field, the temperature does not exceed $2.2$~K. This is the major experimental finding of this work.

It is interesting to compare our finding with the flame temperature derived from magnetic deflagration theory for Fe$_8$. The theory of deflagration~\cite{Garanin2007,book2013} relates the propagation velocity of the spin reversal front $v_f$ to the heat conductivity $\kappa$, the barrier height $U$, and flame temperature $T_f$. The relation is:
\begin{equation}
{v_f}(H,T_f) = \sqrt{\frac{\kappa(T_f)}{\tau_0}}exp \left(\frac{-U(H)}{2k_BT_f}\right)
\label{eq:Deflagration}.
\end{equation}
In Fe$_8$, only three transitions are observed. Therefore the effective barrier height is taken as the energy difference between $m=-10$ to $m=-7$ for the matching field of the $n=1$ transition where deflagration was found, namely $U=10$~K. Additionally, $v\sim1$ $m/s$ and $\kappa\sim2\times 10^{-6}   $ $m^2/s$ are known from previous measurements on particular samples that happen to show deflagration~\cite{LeviantDeflagration}. This gives~$T_f=2.5$~K, in good agreement with our measurement. 

Since the energy differences between the metastable ground state ($m=-10$) and the first and second excited states ($m=-9, -8$) for the $n=1$ and $n=2$ transitions are $\approx5$~K and $\approx10$~K, i.e. greater than $2.2$~K, we can assume that less than 10\% of the spins are excited out of the metastable ground state during the $n=1$ tunneling event. Moreover, not all these excited spins flip. Therefore, the LZS theory should work well for these $n>0$ values even for large crystals. However, it does not. The problem is that one cannot account for the tunneling probability as a function of sweep rate with one tunnel splitting value.

\begin{figure}[tbph]
	\begin{center}
		\includegraphics[trim=0cm 0cm 0cm 0cm ,clip=true,width=8cm]{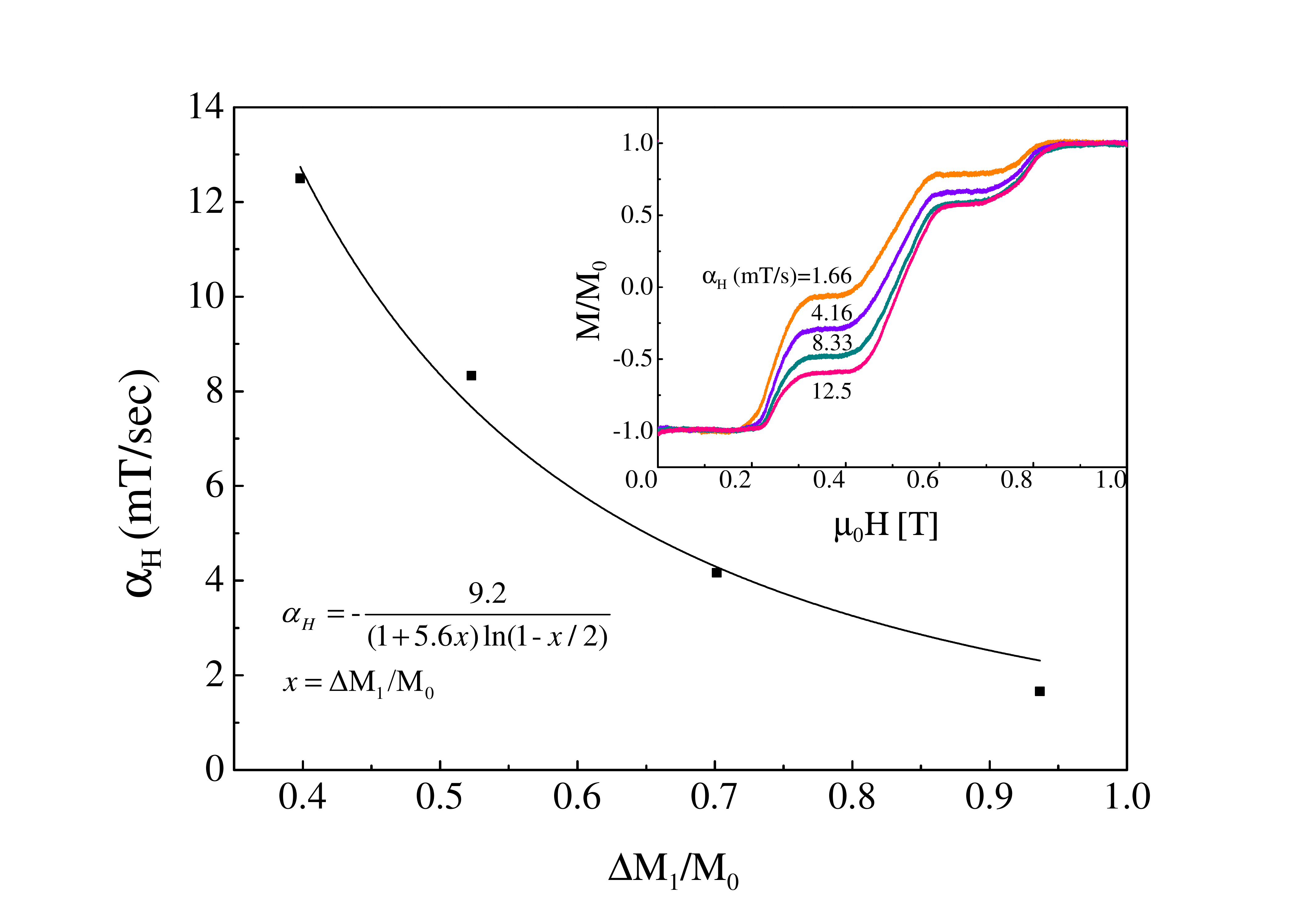}
	\end{center}
	\caption{The magnetic field sweep rate plotted as a function of the first magnetization jump $n=1$ of Fe$_8$. The solid line is a fit to Eq.~\ref{eq:Yaari} derived in the text. The inset is raw hysteresis loop data taken from Ref.~\cite{LeviantDeflagration} from which the magnetization jumps are derived.}
	\label{fig:TunnSplit}
\end{figure}

For the $n=1$ jump we suggest the following explanation for the discrepancy between the LZS theory and the experimental result. The magnetic induction experienced by the spins in the $z$ direction is given by $B=\mu_0 (H+M)$, where $H$ is the magnetic field and $M$ is the uniform sample magnetization. As we sweep $H$ through a transition, $B$ changes according to $\frac{dB}{dt}=\alpha_H+\alpha_M$ where now only $\alpha_H=\mu_0 \frac{dH}{dt}$ and $\alpha_M=\mu_0 \frac{dM}{dH}\frac{dH}{dt}$. We approximate $\frac{dM}{dH}$ by $f (\Delta M/M_0)(M_0/\Delta H)$ where $\Delta H$ is the field width during which the transition is taking place, $\Delta M$ is the magnetization jump, $M_0$ is the saturation magnetization, and $f$ is on the order of unity. The reason for introducing the factor $f$ is that the local variation in the magnetization could be larger than the global one estimated from $\Delta M/\Delta H$. $\Delta H$ is independent of sweep rates, therefore we absorb $M_0/\Delta H$ into $f$ and write
\begin{equation}
\frac{dB}{dt}=\alpha_B=\alpha_H(1+f \frac{\Delta M}{M_0} ).
\label{Lev}
\end{equation}
For the first transition $\Delta M_1=2 M_0 P_1$ where $\Delta M_1$ and $P_1$ stand for the magnetization jump and transition probability at $n=1$, respectively. Substituting Eq.~\ref{Lev} into Eq.~\ref{eq:LZ} and solving for $\alpha_H$ one finds the relation 
\begin{equation}
\alpha_H=\frac{-\pi \Delta_1^2}{2g\mu_B(19)(1+f\frac{\Delta M_1}{M_0})\ln(1+\frac{\Delta M_1}{2M_0})}.
\label{eq:Yaari}
\end{equation}

In Fig.~\ref{fig:TunnSplit} we present the sweep rates as a function of the normalized magnetization jumps for $n=1$. The inset shows raw data taken from Leviant work~\cite{LeviantPhotons}. Due to the limited number of data points the fit parameters are determined with large error bars. We therefore only demonstrate here that there is a set of parameters which capture the data points reasonably well, and estimate the value of the tunnel splitting roughly to be $\Delta_{-10,9}\approx26.6\cdot10^{-7}$K, which is in good agreement with previous work~\cite{Wernsdorfer2000}. The factor $f$ is indeed on the order of unity.

As for the $n \geq 2$ magnetization jumps: They become smaller as the sweep rate decreases. This is contrary to the expectation from the LZS theory. An enhancement of the sweep rate due to the magnetization reversal, as our model implies, could only make the situation worse. Therefore, a description of the high $n$ magnetization jumps is outside of the scope of the LZS theory. We speculate that when the $n\geq 2$ transitions are taking place, the temperature increase of the crystal is enough to excite a substantial portion of the spins to levels from which they flip with high enough probability in a classical manner.

To summarize, we found that when the spin of an Fe$_8$ molecular magnet tunnels from a metastable ground state to an excited state and from there to the stable ground state, the temperature does not increase above $2.2$~K. This suggests that the LZS theory should be valid for large crystals at all magnetization jumps although the sample warms up during the $n\geq 1$ tunneling events. We show that by considering the effective sweep rate, which is affected by the magnetization reversal itself, rather than the external magnetic field sweep rate, we can reach a reasonable agreement between measurements and the LZS theory, but only for the first magnetization jump.

\subsection*{Acknowledgment}

We thank Lev Melnikovsky for helpful discussions. This study was partially supported by the Russell Berrie Nanotechnology Institute, Technion, Israel Institute of Technology.


\end{document}